\documentclass[useAMS,usenatbib]{mn2e}
\usepackage{graphicx}
\usepackage{amssymb}
\usepackage{amsmath}

\newcommand{\B}[1]{\mathbf{#1}}
\newcommand{\Br}{\B{r}}
\newcommand{\Bv}{\B{v}}

\newcommand{\Eq}[1]{Eq.~(\ref{#1})} 
\newcommand{\Fig}[1]{Fig.~\ref{#1}}
\newcommand{\Sec}[1]{Sec.~\ref{#1}}
\newcommand{\EqDef}{\equiv}
\newcommand{\tw}{\tilde{\omega}}

\title{A numerical comparison of theories of violent
  relaxation}

\author[I.~Arad and P.H.~Johansson]{I.~Arad\thanks{Email:
    arad@ast.cam.ac.uk} and
  P.H.~Johansson\thanks{Email: phjohans@ast.cam.ac.uk}\\
  Institute of Astronomy, Madingley Road, Cambridge CB3 OHA, UK}
\begin{document}

\date{Version of \today}
\pagerange{\pageref{firstpage}--\pageref{lastpage}} \pubyear{2004}

\maketitle

\label{firstpage}

\begin{abstract}
  Using $N$-body simulations with a large set of massless test
  particles we compare the predictions of two theories of violent
  relaxation, the well known Lynden-Bell theory and the more recent
  theory by Nakamura. We derive ``weakened'' versions of both theories
  in which we use the whole equilibrium coarse-grained distribution
  function $\bar{f}_i$ as a constraint instead of the total energy
  constraint.  We use these weakened theories to construct expressions
  for the conditional probability $K_i(\tau)$ that a test particle
  initially at the phase-space coordinate $\tau$ would end-up in the
  $i$'th macro-cell at equilibrium.  We show that the logarithm of the
  ratio $R_{ij}(\tau) \EqDef K_i(\tau)/K_j(\tau)$ is directly
  proportional to the initial phase-space density $f_0(\tau)$ for the
  Lynden-Bell theory and inversely proportional to $f_0(\tau)$ for the
  Nakamura theory.  We then measure $R_{ij}(\tau)$ using a set of
  $N$-body simulations of a system undergoing a gravitational collapse
  to check the validity of the two theories of violent relaxation. We
  find that both theories are at odds with the numerical results, both
  qualitatively and quantitatively.
\end{abstract}



\section{Introduction}

A statistical theory that successfully describes the process of a
collisionless gravitational collapse has been a longstanding open
problem. The ultimate goal of such a theory is to predict the (semi-)
equilibrium state of a gravitational system that is described by the
Vlasov equation, given its initial phase-space density $f_0(\Br,\Bv)$.
The implications of such a theory to cosmology and astrophysics are
immense. For example, such a theory may be able to explain the origin
of cusps found in simulations of CDM haloes, the universality of their
shape, and possibly connect them to the initial cold phase-space
density distribution.

The first attempt to construct such a theory was made by Lynden-Bell
in his pioneering work from 1967 \citep[hereafter the LB67
theory]{ref:Lyn67} which actually coined the phrase ``violent
relaxation'' to describe the relaxation of such systems. However,
already in that paper Lynden-Bell predicted that his theory will not
be applicable to large parts of the system due to the ``incompleteness
of violent relaxation''. The fluctuations of the gravitational
potential, which drive the system towards an equilibrium state fade
out too soon for the system to explore the full configuration space.
Therefore the final equilibrium state picked by the system is not
necessarily the one that maximises the entropy. The discrepancy
becomes larger as we go from the centre of the system to its outer
parts where the gravitational fluctuations are much smaller.
Accordingly, Lynden-Bell assumed that the complete relaxation is
limited to a sphere of radius $R_1$ around the system's centre of
mass.

Soon after this paper, many $N$-body simulations were run to check its
validity. Most of these simulations were of one-dimensional models of
either plane sheets or of spherical shells \citep{ref:Coh68,
  ref:Hen68, ref:Cup69, ref:Gol69, ref:Lec71, ref:Tan87}. In most
cases the LB67 theory was only partially correct at best. A common
outcome of the ``water-bucket'' initial conditions (in which there is
only one level of phase-space density surrounded by a vacuum) was the
``core-halo'' structure in which the halo took most of the energy of
the system, leaving the core (which contained most of the mass)
degenerate. Compared with predictions, the phase-space density was too
high at very low energies, too low at intermediate energies, and
oscillated at high energies.  One can find a good fit for the core
using the degenerate solution of the LB67 theory, and a reasonable fit
for the halo using the non-degenerate solution - but then it is
difficult to find a clean way to decide which particles should be
described by which one of the two solutions.  Consequently, the theory
loses much of its predictive power. This ambiguity was attributed to
the incompleteness of the relaxation: the inner parts where the
relaxation was effective could be described by a LB67 solution, unlike
the outer parts which were only partially relaxed.

The disagreement between the predictions of the LB67 theory and
experiments, in addition to some disturbing conceptual issues it
possesses (such as its infinite mass prediction in 3D and phase-space
densities segregation - see \Sec{sec:probs} for further details), have
led people to consider alternatives to it, e.g., \citet{ref:Shu78,
  ref:Sti87, ref:Spe92, ref:Kul97, ref:Nak00, ref:Tre05}. In all these
theories the equilibrium state is assumed to be the most probable
state of the system, which is found by maximising the entropy under
the appropriate constraints. The difference between these theories is
mainly due to the definition of entropy they use and the constraints
under which it is maximised.

For the purpose of this paper we roughly divide these theories into
two groups: in the first group we have theories which are based on a
more fundamental approach to the problem, thereby incorporating
(almost) all of the dynamical constraints when maximising the entropy.
In this group we essentially have two theories: the theory of
\citet{ref:Lyn67} and the theories of \citet{ref:Kul97} and
\citet{ref:Nak00}, which as we shall see, are basically the same
theory. In the second group we have theories that use more heuristic,
ad-hoc like, constraints and entropy definitions, such as
\citet{ref:Sti87, ref:Spe92, ref:Tre05}, or \citet{ref:Han04} who use
non-extensive statistical mechanics, and to some extent
\citet{ref:Shu78}.  Non of these theories, for example, take into
account the conservation of the phase-space volume by the dynamics, as
required by Liouville's theorem. As a rule of thumb, the predictions
of the second group are in a better agreement with simulations and
observations than the first group.  However, this is achieved at the
expense of increasing the number of assumptions and free parameters in
the theory. In this paper we are interested in examining how well the
fundamental assumptions of statistical mechanics apply in violent
relaxation, we shall therefore concentrate on the first group of
theories: the LB67 theory and the theory of \citet{ref:Nak00}
(hereafter NK00).

The purpose of this paper is three-fold: firstly, we wish to compare
the predictions of the LB67 and NK00 theories with numerical
experiments \emph{done in 3D}. To our knowledge this has not yet been
done directly, in particular when it comes to the NK00 theory.  While
the 1D case shares many common characteristics with the 3D case, it is
known that dimensionality may play an important role in systems
undergoing violent relaxation.  For example, the 3D system has more
degrees of freedom, and may therefore experience a more efficient
mixing.  Indirect support for the NK00 theory in $3D$ is found in
\citet{ref:Mer03}. Here the authors show that the velocity
distribution function of isolated systems after violent relaxation is
well fitted by a single Gaussian, as is predicted by the NK00 theory
for the non-degenerate case. Nevertheless, we would like to perform a
more direct test of the predictions of this theory.

Secondly, acknowledging the fact that these theories are largely
incorrect, we would still like to see if certain aspects of them are
true. In other words, we would like to see if the approach of
maximising the entropy under the proper constraints is at all useful
\emph{in the weakest possible sense} for a dissipationless
self-gravitating system. Thirdly, as we shall see in \Sec{sec:theory},
the NK00 theory and the LB67 theory have a very similar structure,
incorporating the same set of constraints, while using fundamentally
different entropy definitions. We would therefore like to know which
definition describes violent relaxation better. We feel that somehow
the question of how entropy should be defined is more fundamental than
the full statistical theory around it. For example, a local definition
of entropy can be used in a dynamical theory that describes the
approach to equilibrium [e.g. \citet{ref:Cha98b}].

The structure of this paper is as follows: in \Sec{sec:theory} we
review the LB67 and NK00 theories and derive them using the
information-theory approach. We introduce the conditional and joint
probabilities that describe the path of a test particle and relate
them to the different definitions of entropy between the two theories.
We then list some of the problems and questions with these theories
and re-derive ``weakened'' versions of the theories. We show how the
predictions of these theories can be tested in a numerical experiment
that to a large extent avoids many of the pitfalls we would have
encountered had we tried to test the LB67 and NK00 theories in their
original form. In \Sec{sec:measure-K} we describe how the conditional
probabilities can be measured in an $N$-body simulation using a large
number of test particles. Then in \Sec{sec:simulations} we describe
the numerical simulations that we have performed to measure these
probabilities and in \Sec{sec:results} we present the results of the simulations.  
Our conclusions are presented in \Sec{sec:conc}.


\section{Theoretical background - two theories for violent relaxation}
\label{sec:theory}

In this section we will focus on the LB67 theory and a more recent
alternative to it - the NK00 theory. The aim of these theories is to
predict the final equilibrium state of a collisionless gravitating
system, given its initial state.  The state of the system is
completely determined by its phase-space density function (DF)
$f(\Br,\Bv,t)$, which is governed by the Vlasov equation
\begin{equation}
  \label{eq:Vlasov}
  \partial_t f + \Bv \cdot \partial_{\Br}f - \nabla \Phi \cdot
  \partial_{\Bv}f = 0 \ .
\end{equation}
Here $\Phi(\Br,t)$ is the gravitational potential,
calculated self-consistently from $\rho(\Br,t) = \int\!\!
f(\Br,\Bv,t)d^3\Bv$ via Poisson's equation. 

\subsection{Derivation of the NK00 theory}

The NK00 theory was originally derived in the framework of information
theory. Here we shall repeat its
derivation for comparison with the LB67 theory. We first formulate the
violent relaxation process as an experiment with a well defined set of
possible results $\omega_i$ and a set of corresponding probabilities
$p_i$. The most probable result is then the one that maximises the
Shannon entropy $S=\sum_i p_i\log p_i$ under some prescribed
constraints on $p_i$ \citep{ref:Jay57a, ref:Jay57b}.

Let $f_0(\tau)$ be the initial phase-space density of the system, with
$\tau\EqDef (\Br,\Bv)$ being a phase-space coordinate. We toss a test
particle into the initial phase-space according to the probability
distribution $p_0(\tau)$ which is defined by
\begin{equation}
  p_0(\tau) = \frac{1}{M}f_0(\tau) \ ,
\end{equation}
and let it move under gravity just like any phase-space element. We
define $p(\tau,t)$ to be the probability distribution of finding the
test particle at time $t>0$. The conservation of phase-space volume
guarantees that $p(\tau,t)=f(\tau,t)/M$ for all $t>0$.

Next, we divide phase-space into macro-cells $i=1,2,3,\ldots$ of
volume $\tw$, and define the coarse-grained probability $\bar{p}_i$ as
the probability of finding the test particle in the $i$'th macro-cell
when the system reaches an equilibrium. From the above discussion it
is clear that $\bar{p}_i$ is equal to $\bar{f}_i/M$ with $\bar{f}_i$
being the coarse-grained DF in the macro-cell $i$ at equilibrium.

A possible result of our experiment is the pair $(\tau,i)$ which
specifies the initial location of the test particle and the number of
the cell where it is finally found. With each such pair we associate
the \emph{joint probability-distribution}
$p_i(\tau)$. $p_i(\tau)d^6\tau$ is the probability that the particle
was initially in a small patch of volume $d^6\tau$ around $\tau$, and
ended up in the $i$'th macro-cell. The Shannon entropy of our
experiment is then
\begin{equation}
  \label{eq:S-NK}
  S_{NK} = -\sum_i\int\!\!d^6\tau \, p_i(\tau)\log p_i(\tau) \ .
\end{equation}
Maximising $S_{NK}$ gives us the most probable $p_i(\tau)$, from which
we can get
\begin{equation}
  \label{eq:coarse-f}
  \bar{f}_i = M\int\!p_i(\tau) d^6\tau \ .
\end{equation}
Before maximising $S_{NK}$, however, we must write down the
constraints on the probabilities $p_i(\tau)$. The first constraint
stems from the initial conditions and from the fact that $p_i(\tau)$
is a joint probability distribution:
\begin{equation}
  \label{eq:initial-con}
  \sum_i p_i(\tau) = p_0(\tau) = \frac{1}{M}f_0(\tau) \ .
\end{equation}
The second constraint comes from the conservation of phase-space
volume under the collisionless dynamics: the overall
initial phase-space volume of all phase-space elements that ended up
in cell $i$ will be exactly $\tw$:
\begin{equation}
  \label{eq:vol-con}
  M\int\!\! d^6\tau \frac{p_i(\tau)}{f_0(\tau)} = \tw \ .
\end{equation}
This constraint guarantees that the NK00 theory does not violate
Liouville's theorem. Finally, from the conservation of energy, the
total energy in $\bar{f}_i$ must be equal to the initial energy of the
system $E$
\begin{equation}
  \label{eq:energy-con}
  E = \sum_i \tw\bar{f}_i\left[\frac{v^2_i}{2} 
   + \frac{G}{2}\sum_j\frac{\tw\bar{f}_j}
         {|\Br_i - \Br_j|}\right] \ ,
\end{equation}
with $(\Br_i, \Bv_i)$ being the phase-space coordinates of the centre
of the $i$'th cell. 

The initial conditions (\ref{eq:initial-con}) and the phase-space
volume constraints (\ref{eq:vol-con}) can be neatly written in terms
of the \emph{conditional probability distribution} $K_i(\tau)$,
defined by
\begin{equation}
  \label{def:K}
  K_i(\tau) \EqDef \frac{p_i(\tau)}{p_0(\tau)} 
          = M\frac{p_i(\tau)}{f_0(\tau)} \ .
\end{equation}
$K_i(\tau)$ is the probability that a test particle which is initially
at $\tau$ will end-up at the $i$'th cell. Then the initial conditions
and volume preservation constraints can be written as
\begin{eqnarray}
  \label{eq:initial-con-K}
  \sum_i K_i(\tau) &=& 1 \ , \\
  \label{eq:vol-con-K}
  \int\!\! d^6\tau K_i(\tau) &=& \tw \ ,
\end{eqnarray}
while the coarse-grained DF is given by
\begin{equation}
  \label{eq:coarse-f-K}
  \bar{f}_i = \int\!\! f_0(\tau) K_i(\tau)\, d^6\tau \ .
\end{equation}
Using the well known technique of Lagrange multipliers it can now be easily
shown that the $K_i(\tau)$ for which $\delta S_{NK} = 0$ is 
\begin{equation}
  \label{eq:K-NK00}
  K_i(\tau) = e^{-\beta\epsilon_i - \delta(\tau) -
    \lambda_i/f_0(\tau)} \ .
\end{equation}
Here $\epsilon_i = v_i^2/2 + \Phi(\Br_i)$ is the specific energy in
the $i$'th cell and $\beta, \delta(\tau), \lambda_i$ are the Lagrange
multipliers of the energy constraint, the initial conditions
constraint and the volume preservation constraint respectively.

\subsection{The LB67 theory: equal-mass discretisation vs. equal-volume 
 discretisation}
\label{sec:mass-vol}

The NK00 theory was originally derived within the information theory
approach which was outlined in the previous section, while the LB67
theory was derived using a combinatorial approach. Their different
predictions, however, are not related to the different frameworks in
which they were derived, but rather to the different definition of
entropy they use. As noted by \citet{ref:Nak00}, the LB67 theory can
be recovered in the framework of information theory if instead of
defining the entropy in terms of the joint-probability $p_i(\tau)$,
one defines it in terms of the conditional probability $K_i(\tau)$:
\begin{equation}
  \label{eq:S-LB}
  S_{LB} = -\sum_i\int\!\!d^6\tau \, K_i(\tau)\log K_i(\tau) \ .
\end{equation}
Indeed, maximising $S_{LB}$ with respect to the
constraints~(\ref{eq:initial-con}-\ref{eq:energy-con}) yields the
expression
\begin{equation}
  \label{eq:K-LB67}
  K_i(\tau) = e^{-\beta f_0(\tau)\epsilon_i - \delta(\tau) -
    \lambda_i} \ ,
\end{equation}
with $\beta, \delta(\tau), \lambda_i$ being the Lagrange multipliers
as in \Eq{eq:K-NK00}. Using the volume preservation
constraint~(\ref{eq:vol-con}) it is easy to express $\lambda_i$ in
terms of the other unknowns, and plugging this into
\Eq{eq:coarse-f-K}, after a trivial re-definition of $\delta$ we
obtain
\begin{equation}
  \bar{f}_i = \frac{\int_{f_0>0} f_0(\tau) 
     e^{-\beta f_0(\tau)\epsilon_i - \delta(\tau)}d^6\tau}
    {1 + \int_{f_0>0} e^{-\beta f_0(\tau')\epsilon_i -
        \delta(\tau')}d^6\tau'} \ ,
\end{equation}
which is identical to the expression for the coarse-grained DF in the
LB67 theory.

On the other hand, \citet{Ara05} have shown that the NK00 theory
can be derived in a combinatorial approach, where the entropy is
defined simply (up to an additive constant) as the logarithm of the
number of micro-states which comply with a given macro-state. While in
the LB67 theory one discretises phase-space by considering elements
of \emph{equal volume}, the NK00 theory is recovered if one considers
elements of \emph{equal mass}.  Therefore we see that using the joint
probability distribution in the information theory approach is
equivalent to equal-mass discretisation, whereas the use of
conditional probability distribution is equivalent to equal-volume
discretisation.  The discretisation of phase-space into cells of equal
mass was already done by \citet{ref:Kul97}. The NK00 theory is
therefore essentially the same as the \citet{ref:Kul97} theory 
and for that reason we only consider one of them in this paper.

The above analogy can also be demonstrated by modifying the
test particle experiment that was used in the previous section to
derive the NK00 theory. Instead of letting $p_0(\tau)$ - the
probability distribution for finding the test particle initially at
$\tau$ - be proportional to $f_0(\tau)$, we can devise an alternative
experiment in which the particle has an equal probability of being
everywhere in phase-space, provided, of course, that the overall
phase-space volume $V$ is finite. In other words, we let
$p^{(LB)}_0(\tau) = 1/V$. As $p^{(LB)}_0(\tau)$ is a constant, the
conditional probability $K^{(LB)}_i(\tau)$ is proportional to the
joint-probability and therefore the usual Shannon entropy \Eq{eq:S-NK}
will yield the LB67 results. The connection to the equal-volume
discretisation versus equal-mass discretisation difference that is
found in the combinatorial approach is now evident: in the LB67
experiment the probability of initially finding the particle in a
given patch of phase-space is proportional to the volume of the patch,
whereas in the NK00 experiment it is proportional to the mass within
that patch.

Before concluding this section we note an important difference between
the resultant conditional probability in the LB67 theory and the NK00
theory. Whereas in the LB67 theory the coupling between the initial
coordinates $\tau$ and the final coordinate $i$ is in the $\beta
f_0(\tau)\epsilon_i$ term in the exponent, in the NK00 theory it is in
the $\lambda_i/f_0(\tau)$ term. This different coupling will be
extensively discussed in the following sections.

\subsection{Empirical and conceptual problems in the theories}
\label{sec:probs}

The statistical theory of violent relaxation (both LB67 and NK00)
suffers from several problems and open questions. In the next two 
sections we briefly describe some of these problems and sketch a numerical
experiment which will enable us to answer some of these questions.

\begin{description}
\item[\emph{1. Incomplete relaxation}] Firstly, and perhaps most
  importantly, the relaxation is never complete - indeed, as was
  already noticed by \citet{ref:Lyn67}, after a few oscillations on
  the dynamical timescale of $(G\bar{\rho})^{-1/2}$ it is over. This
  does not give the system enough time to probe the configuration
  space, and as a result the system will settle down in a state that
  does not necessarily maximise the entropy.

  Nevertheless, it has often been argued that in the central regions
  of the collapsed object, where $\bar{\rho}$ is high and the
  dynamical time scales are short, violent relaxation is efficient and
  we may therefore expect the system to approach the equilibrium
  solution in these regions. However, it is not clear how this claim
  can be verified quantitatively. The equilibrium state depends on
  the gravitational potential, which in turn depends on the
  phase-space density also in the outer parts of the system, where
  the relaxation is incomplete.

\item[\emph{2. Definition of entropy}] As demonstrated in the last two
  sections, there exist (at least) two equally plausible ways of
  defining the entropy. These two definitions yield two different
  predictions for the equilibrium state, and therefore at least one of
  them is wrong. It would be interesting to check numerically which
  entropy (if either) better describes violent relaxation.

\item[\emph{3. Additional hidden constraints}] It is possible that in the
  violent relaxation process there exist a set of preserved quantities
  that are not taken into account in the theories when maximising
  the entropy. In such a case, the actual configuration space of the
  system is much more limited and its maximal entropy state may be
  different from the calculated one. Moreover, this state would have a
  smaller entropy than the maximal entropy which is calculated without
  the constraints. Such a mechanism may prove to be another reason for
  the incomplete violent relaxation.

  One such uncounted constraint is the total angular momentum of the
  system, which has not been taken into account in the present
  derivations of the LB67 and NK00 theories. In principle, however, it
  can be taken into account by adding the appropriate Lagrange
  multipliers, with the cost of adding some additional complication to the
  final result. Another such invariant is 
  \begin{equation}
    \label{eq:Q}
    Q_\nu \EqDef \int\!\! \bar{f}(\tau)
     \left(\frac{|\B{J(\tau)}|}{\epsilon^{3/4}(\tau)}\right)^\nu
  \end{equation}
  which was suggested by \citet{ref:Sti87} to be \emph{approximately}
  conserved upon a phenomenological basis [see also
  \citet{ref:Tre05}]. Further examples of invariants and their
  influence on the dynamics can be found in \citet{ref:Mou95,
    ref:Hen04}.

  Finally the Hamiltonian nature of the phase-space flow gives rise to a 
  set of constraints that are more difficult to handle. Such flows which can
  always be described by a canonical transformation $(\Br,\Bv) \to
  (\Br',\Bv')$, necessarily preserve the so called \emph{Poincar\'{e}
    integral invariants} which are an infinite set of invariants. The 
  circulation-like integrals which were already mentioned in 
  \citet{ref:Lyn67} are a subset of these invariants. The inclusion of 
  such invariants in a statistical theory of violent relaxation is a 
  mathematical challenge, which to the best of our knowledge is yet 
  to be overcome.

\end{description}

\subsection{The weakened versions of LB67 and NK00}
\label{sec:num-exp}

To study these three points we have devised a numerical experiment in
which the conditional probability $K_i(\tau)$ is directly measured. It
is then compared to weakened versions of the LB67 and NK00 theories in
which the \emph{whole equilibrium coarse-grained DF $\bar{f}_i$ is
  taken as a constraint}, thereby replacing the total energy
constraint.  To see how this construction may shed light on the above
points, let us first derive the weakened versions of the LB67 and NK00
theories.

We start with the NK00 theory, replacing the energy
constraint~(\ref{eq:energy-con}) with the ``final conditions''
\Eq{eq:coarse-f-K}, which is now taken as a constraint. The Lagrange
multiplier $\beta$ is thus replaced with a set of multipliers $\xi_i$,
and the functional we wish to maximise is

\begin{eqnarray}
\label{eq:I-nk}
   I_{NK} &=& -\sum_i\int\!\!d^6\tau \,
     f_0(\tau)K_i(\tau)\ln[f_0(\tau)K_i(\tau)] \nonumber \\
    &&+\ \sum_i\int\!\!\Big[ \delta(\tau)K_i(\tau) 
    + \lambda_iK_i(\tau) \nonumber \\
    &&+\ \xi_if_0(\tau)K_i(\tau)\Big]\,d\tau \ .
\end{eqnarray}
The functional derivative of $I$ with respect to $K_i(\tau)$ is then
\begin{eqnarray}
  \frac{\delta I_{NK}}{\delta K_i(\tau)} 
   &=& -f_0(\tau)\big\{\ln [f_0(\tau)K_i(\tau)] + 1\big\} +
   \delta(\tau)  \nonumber \\
   &&+\ \lambda_i + f_0(\tau)\xi_i \ ,
\end{eqnarray}
and equating it to zero we obtain
\begin{equation}
  \ln K_i(\tau) = \frac{\delta(\tau)}{f_0(\tau)} + \xi_i 
     + \frac{\lambda_i}{f_0(\tau)} \ ,
\end{equation}
after trivial redefinitions of $\delta(\tau)$ and $\xi_i$.

To derive the equivalent result for the LB67 theory, we replace
$f_0(\tau)K_i(\tau) \to K_i(\tau)$ in the definition of the entropy in
\Eq{eq:I-nk}, while leaving the constraints intact.  We thus obtain a
new functional $I_{LB}$. Differentiating it with respect to
$K_i(\tau)$ we get
\begin{equation}
  \frac{\delta I_{LB}}{\delta K_i(\tau)} 
   = -\ln K_i(\tau) + 1 + \delta(\tau) + \lambda_i 
     + f_0(\tau)\xi_i \ ,
\end{equation}
and therefore the extremum solution is 
\begin{equation}
  \ln K_i(\tau) = \delta(\tau) + \lambda_i 
     + f_0(\tau)\xi_i \ ,
\end{equation}
after redefining $\lambda_i \to \lambda_i+1$.  As noticed at the end
of section \ref{sec:mass-vol}, the most striking difference between the two
theories is the coupling between the initial coordinate $\tau$ and the
final coordinate $i$. This can be summarised as follows
\begin{eqnarray}
\label{eq:K-NK}
  K^{(NK)}_i(\tau) &=& A^{(NK)}(\tau)\, 
       B_i^{(NK)}\, e^{\frac{c_i^{(NK)}}{f_0(\tau)}} \ , \\
\label{eq:K-LB}
  K^{(LB)}_i(\tau) &=& A^{(LB)}(\tau)\, 
       B_i^{(LB)}\, e^{c_i^{(LB)}f_0(\tau)} \ .
\end{eqnarray}
We see very prominent differences between the two theories, which may
be detectable numerically if one measures $K_i(\tau)$ directly.
However, to find the theoretical predictions for $A(\tau), B_i$ and
$c_i$ one has to solve the constraints Eqs.~(\ref{eq:initial-con-K},
\ref{eq:vol-con-K}, \ref{eq:coarse-f-K}), which is a non-trivial task.
This can be largely avoided if we focus on the \emph{ratio}
\begin{equation}
\label{eq:Rij}
  R_{ij}(\tau) \EqDef K_i(\tau)/K_j(\tau)
\end{equation}
for some fixed $i,j$ coordinates. The predictions of the NK00 and LB67
theories for this quantity are:
\begin{eqnarray}
  R^{(NK)}_{ij}(\tau) &=& 
      \frac{B_i^{(NK)}}{B_j^{(NK)}} e^{\frac{c_i^{(NK)}-c_j^{(NK)}}{f_0(\tau)}} \ , \\
  R^{(LB)}_{ij}(\tau) &=& 
      \frac{B_i^{(LB)}}{B_j^{(LB)}} e^{\left(c_i^{(LB)}-c_j^{(LB)}\right) f_0(\tau)} \ .
\end{eqnarray}
We notice that in both cases the only dependence of $R_{ij}(\tau)$ on
$\tau$ is through $f_0(\tau)$, and therefore by plotting $R_{ij}$ as a
function of $f_0(\tau)$ we can see if either of the theories hold.
Specifically, the LB67 theory predicts that $\ln R_{ij}(\tau)$ is
linear in $f_0(\tau)$ whereas the NK00 theory predicts that it is
linear in $1/f_0(\tau)$. This is the central idea of this paper.

This is of course only a sufficient test. By passing it, we are not
guaranteed that $K_i(\tau)$ is given by \Eq{eq:K-NK} or \Eq{eq:K-LB}.
Moreover, even if this is the case, it is still not the form of the
original NK00 and LB67 theories which use the much weaker energy
constraint instead of the full coarse-grained $\bar{f}_i$ constraint.
But this is also the advantage of using such a test: we already know
that these theories are wrong to a large extent, in particular at the
outer parts of the collapsed object.  Therefore just because this test
is a very weak test, it allows us to see if there is a little grain of
truth in either of them. If neither of the theories passes the test
then there is an extremely small chance that the idea of maximising
the entropy, at least as defined in the NK00 and LB67 theories, has
any physical relevance. On the other hand, if one of these theories
passes the test it would allow us to decide what is the preferable
form of the entropy.

The proposed test also partly circumvents the difficulties in points 1
and 3: firstly, by picking $i$, $j$ and $\tau$ to be well inside the
collapsed object, we are assured that the region of phase-space that
we are measuring has undergone the maximal mixing which the simulation
provides. Of course it does not mean that this is the maximal mixing
that is necessary for the entropy to reach its predicted maximum - but
it is ``as good as it can get''. 

Secondly, as the above predictions for $R_{ij}(\tau)$ were made on the
basis of theories which use the final $\bar{f}_i$ as a constraint, any
failure of the theories in passing the test could not be attributed to
any unknown constraint which involve $\bar{f}_i$ (such as energy and
total angular momentum).  In other words, the inclusion of such
constraints would not change the predictions of the theories for
$R_{ij}(\tau)$. Other constraints, which involve $K_i(\tau)$ itself,
such as possibly the constraints due to the conservation of
circulation integrals, may still affect the predictions.

\section{Measuring conditional probabilities in an $N$-body
  simulation}
\label{sec:measure-K}

To test the predictions of the LB67 and NK00 theories we ran a set of
$N$-body simulations of a gravitational collapse. The goal of these
simulations was to measure the conditional probability $K_i(\tau)$
from which the ratio $R_{ij}(\tau)=K_i(\tau)/K_j(\tau)$ can be
calculated. In what follows we explain how this probability is
measured.

\subsection{Using test particles to measure $K_i(\tau)$}
\label{sec:testp}

Theoretically, the conditional probabilities $K_i(\tau)$ can be
measured in an $N$-body simulation using a large set of $N_t$ test
particles which are placed initially in a small patch of phase-space
around $\tau$. The gravitational mass of the test particles is set to zero so that
they do not affect the evolution of the system - but only trace it.
Then when the system relaxes one can estimate the phase-space density
of the test particles by re-assigning to them a mass of $m_t = 1/N_t$,
such that their total mass is unity. The resultant density in the
macro-cell $i$ is then simply $K_i(\tau)$.

In practise, however, this procedure might fail as it is well known
that in almost any $N$-body simulation the trajectory of each particle
exponentially departs from the exact trajectory of the Vlasov equation
due to chaos \citep[see for example][]{ref:Heg91, ref:Qui92}.  These
deviations, however, may only have a small effect on
\emph{statistical} measurements like the average density of some
region in space, since the errors of individual particles tend to
cancel out each other.  On the other hand, this cancellation, is not
guaranteed when considering a group of test particles which are
initially located in a tiny patch of phase-space. Such a configuration
may be very sensitive to the particular realisation of the massive
particles since all the trajectories of the test particles are very
close to each other and they can all be influenced simultaneously by a
single massive particle. As a result, two simulations that use exactly
the same initial phase-space density but with different realisations
(say, by choosing a different random-number-generator seed) would
produce significantly different results when measuring the test
particles distribution at some final time $t$. The distribution of the
massive particles, however, would be almost statistically identical.
This behaviour was observed when we first tried to measure $K_i(\tau)$
using the above method.

To overcome this difficulty we spread the test particles over large 5D
regions in phase-space in which $f_0(\Br,\Bv)$ is constant. We denote
these regions by $V_0$.  The assumption here is that in such regions
the diversity in the test particles trajectories is large enough to
produce stable statistical results.  This of course has to be checked
numerically by running the same simulation with a different
realisation of the massive particles.

Finally, a similar situation exists in the final snapshot, when one
wishes to recover the phase-space density of the test particles.  A
single-point measurement tends to fluctuate both in time and as a
function of the initial realisation due to Poissonian noise. These
fluctuations can be removed by measuring the \emph{average}
phase-space density on surfaces over which we assume from symmetry
considerations that the exact phase-space density (i.e., the exact
solution of the Vlasov equation) is constant.  For example, if the
equilibrium system is spherical and isotropic, we can choose the
surface to be a sphere around its centre with $\Bv=0$.  We denote the
final surfaces by $S_1$. Consequently, the averaged conditional
probability over the initial region $V_0$ and the final surface $S_1$
is denoted by $\left<K(S_1, V_0)\right>$.

Even after averaging over $V_0$ and $S_1$ our results were sensitive
to the particular realisation of the massive particles that was used.
To estimate this sensitivity we measured $\left<K(S_1, V_0)\right>$
from five simulations with different realisations of the initial
conditions, and calculated the average and scatter. As discussed in
\Sec{sec:results}, the typical scatter between the different
realisations was of the order of 20\%-30\%.

We conclude this section by noting that the averaging procedure over
$V_0$ and $S_1$ does not change the prediction of the theories. To see
why this is the case, let us pass to a continuous description of the
conditional probabilities by replacing the macro-cell coordinate $i$
with the continuous coordinate $\bar{\tau}$ which specifies the
phase-space coordinate of the centre of the macro-cell $i$. Then the
conditional probability is defined such that $K(\bar{\tau},
\tau)d^6\bar{\tau}$ is the probability that a particle that was
initially at $\tau$ will end up in a small phase-space region
$d^6\bar{\tau}$ around $\bar{\tau}$.  The predictions of the
generalised NK00 and LB67 theories for the conditional probabilities,
which are given in Eqs.~(\ref{eq:K-NK}, \ref{eq:K-LB}), can now be
written as
\begin{eqnarray}
\label{eq:K-NK-cont}
  K^{(NK)}(\bar{\tau},\tau) &=& A^{(NK)}(\tau)\, 
      B^{(NK)}(\bar{\tau})\, 
      e^{\frac{c^{(NK)}(\bar{\tau})}{f_0(\tau)}} \ ,
  \\
\label{eq:K-LB-cont}
  K^{(LB)}(\bar{\tau}, \tau) &=& A^{(LB)}(\tau)\, 
      B^{(LB)}(\bar{\tau})\, 
      e^{c^{(LB)}(\bar{\tau})f_0(\tau)} \ . 
\end{eqnarray}
Then the average conditional probability over $V_0$ and $S_1$ is
\begin{eqnarray}
\nonumber
  \left<K(S_1,V_0)\right> &\EqDef& |V_0|^{-1}|S_1|^{-1}
       \int_{V_0} d^5\tau' \int_{S_1} d^2\bar{\tau}'\,
       K(\bar{\tau}', \tau') \\
   &=& |V_0|^{-1} \int_{V_0} d^5\tau'\,
       K(\bar{\tau}', \tau') \ .
\end{eqnarray}
In the last equality we used the assumption that $K(\bar{\tau},\tau)$
is identical for all $\bar{\tau}\in S_1$.  Plugging the expressions
for $K^{(NK)}(\bar{\tau},\tau)$ and $K^{(LB)}(\bar{\tau},\tau)$ into
the above equation, and using the fact that $f_0(\tau)$ is constant
over $V_0$, we get
\begin{eqnarray}
  &&\left<K^{(NK)}(S_1,V_0)\right> \nonumber \\
    &&\quad = 
\label{eq:K-NK-av}
      \left<A^{(NK)}(\tau)\right>_0\, B^{(NK)}(\bar{\tau})\, 
       e^{\frac{c^{(NK)}(\bar{\tau})}{f_0}} \, \\
  &&\left<K^{(LB)}(S_1, V_0)\right>  \nonumber \\
\label{eq:K-LB-av}
      &&\quad = \left<A^{(LB)}(\tau)\right>_0\, B^{(LB)}(\bar{\tau})\, 
  e^{c^{(LB)}(\bar{\tau})f_0} \ ,
\end{eqnarray}
with $\left<\cdot\right>_0$ denoting an average over the initial
region $V_0$. 

We thus see that the structure of the conditional probabilities in
both theories remains unchanged.

\subsection{Numerically estimating the phase-space density of the
  test particles}
\label{sec:esti}

Estimating the 6D phase-space density of few tens of thousands
particles is by no means a trivial task. A simple box-counting
procedure with equal volume boxes is impractical as the number of
boxes in the 6D phase-space overwhelmingly exceeds the number of
available particles, unless we choose the boxes to be so large that
the resultant resolution is extremely poor. The solution is to use an
adaptive technique such as the kernel-based technique in SPH
simulations. In this work, we used the Delaunay Tessellation Field
Estimator (DTFE) method which was introduced by \citet{ref:Sch00} to
estimate real space densities, and was later adapted by
\citet{ref:Ara04a} for the estimation of phase-space densities. To
calculate the average phase-space density on the $S_1$ sphere we used
a Monte-Carlo integration technique by randomly picking $N_{MC}=5,000$
points on $S_1$ and calculating the average phase-space density at
these points. \citet{ref:Ara04a} demonstrated that
$f^{DTFE}/f^{exact}$ is approximately given by a log-normal
distribution, and therefore we calculated the average $K$ using a
geometrical mean:
\begin{equation}
\label{eq:DTFE-average}
  \log \left<K(S_1, V_0)\right> \EqDef \frac{1}{N_{MC}}
           \sum_{j=1}^{N_{MC}} \log[f_t^{DTFE}(\tau_j)] \ . 
\end{equation}
In the above formula $\tau_1, \ldots, \tau_{N_{MC}}$ are the
Monte-Carlo sampling points on $S_1$ and $f_t^{DTFE}(\tau_j)$ is the
DTFE estimate for the phase-space density of the test particles at
$\tau_j$ - test particles that were placed initially in the region
$V_0$.

\begin{figure}
  \center{\includegraphics[scale=0.9]{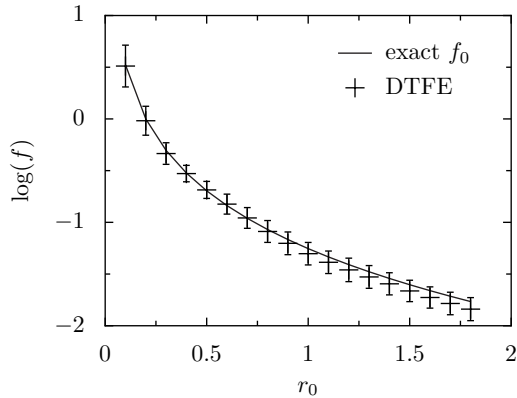}}
  \caption{The DTFE-measured average phase-space density of the
    massive particles on surfaces where the exact phase-space density
    is constant. Each average phase-space density was calculated using
    5,000 random points on the surface. The error-bars denote the
    Poissonian error which is typically 30\%. However, the error in
    the \emph{average} density itself is much smaller, typically about
    7\% and it never exceeds 10\%.} 
  \label{fig:check-DTFE}
\end{figure}

To estimate the internal DTFE measurement error we used the same
Monte-Carlo technique to estimate the phase-space density in a
synthetic distribution realised using $100,000$ particles (the same
number of test particles as was used in the simulations, see next
section). The $f(\Br,\Bv)$ of the synthetic distribution was identical
to the initial distribution of the massive particle in the experiment
(see next section). This is a Hernquist-like distribution with
Gaussian velocity distribution, described by
Eqs.~(\ref{eq:initial-f}-\ref{eq:initial-sigma}). We estimated the
phase-space density on spheres with $0.1<r<1.8$ and $\Bv=0$.  The
results are presented in \Fig{fig:check-DTFE}.  We see that the DTFE
method produces a Poissonian scatter of about 20\%-30\% around its
mean, which in turn matches the exact phase-space density within an
error of no more than 10\%.

We notice that the average DTFE values are systematically smaller than
the exact density.  This can be explained by the fact that we are
measuring the phase-space density on a surface where $\Bv=0$.
According to \Eq{eq:initial-f} in the next section, such a surface
corresponds to a local maxima in $f_0(\Br,\Bv)$ and since the density
in a given point is calculated by linearly interpolating the
phase-space density of nearby particles, we expect it to be lower than
the exact value.  Indeed for smaller radii, where the mean separation
between particles is smaller, the systematic deviation is smaller.
Such systematic errors, however, will be largely cancelled out from
the ratio in \Eq{eq:Rij}.

\section{Numerical Simulations}
\label{sec:simulations}

We used a large set of numerical simulations to measure $\left< K(S_1,
  V_0)\right>$ for 12 initial regions $V_0$ and 5 final surfaces
$S_1$. All simulations used the same initial phase-space density
$f_0(\Br,\Bv)$ which was realised using 120,000 massive particles.
Each of the initial $V_0$ regions was sampled using 100,000 test
particles. The simulations were run in physical coordinates with no
cosmological expansion and in dimensionless mode with the
gravitational constant, unitlength, unitmass and unitvelocity all set
to unity.

As was mentioned in \Sec{sec:testp}, we used 5 different realisation
of essentially the same simulation in order to estimate the
sensitivity of the final $\left<K(S_1,V_0)\right>$ to a particular
realisation. Therefore, for each $V_0$ we run a set of 5 different
simulations in which we used a different realisation of $f_0(\Br,\Bv)$
(by using a different random-number-generator seed in the routine that
set the initial positions and velocities of the massive particles).
The initial positions and velocities of the test particles in $V_0$
were always the same. Our results, which are presented in
\Sec{sec:results}, are based on the average of these different
realisations.

In the following sections we give a detailed description of the
initial conditions and of the numerical simulations themselves.

\subsection{Initial conditions}
\label{sec:initial}

\subsubsection{Massive particles}
\begin{figure}
 \begin{center}
   {\includegraphics[scale=0.32]{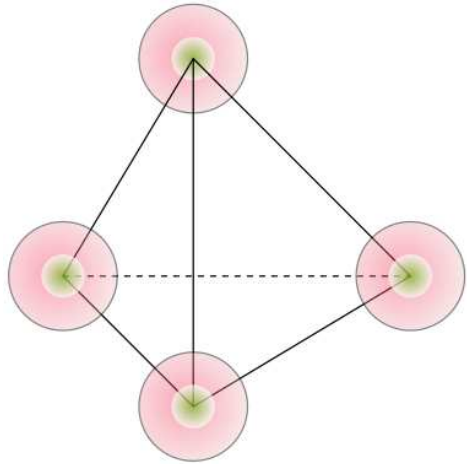}}
 \end{center}
 \caption[]{An illustration of our initial conditions showing the
   four haloes arranged in a symmetrical tetrahedron. The figure
   depicts the massive particles embedded with the massless test
   particles in the centre of each halo.} 
 \label{fig:violent-init}
\end{figure}

Initially, the system consisted of four identical haloes that collapse
into each other, producing a single virialised halo.  Each halo was
realised using 30,000 massive particles, giving a total of 120,000
massive particles in each simulation.  The haloes were placed in a
symmetrical tetrahedron around the origin with their respective
centres separated by six length units.  This setup was chosen to
maximise the level of violent relaxation and to ensure that there was
no preferred direction in the final merger of the haloes.  The test
particles were then placed in equi-density regions inside each halo,
as explained in the next section. A schematic picture of this setup is
shown in \Fig{fig:violent-init}: the four haloes are represented by
the large spheres while the test particles are represented by the
smaller spheres which are embedded inside them.

The haloes were set to be spherical and isotropic with Gaussian
velocities:
\begin{equation}
\label{eq:initial-f}
   f_0(\Br, \Bv) = (2 \pi)^{-3/2}\frac{\rho(r)}{\sigma^{3}(r)} 
   e^{-\frac{v^{2}}{2\sigma^{2}(r)}}. 
\end{equation}
We used a Hernquist-like density profile
\begin{equation}
\label{eq:initial-rho}
   \rho(r)=\frac{1}{r(1+r)^3} \ ,
\end{equation}
while the velocity dispersion $\sigma(r)$ was set by the Jeans
equation \citep{ref:Bin87} with a vanishing anisotropy parameter
$\beta$
\begin{equation}
\label{eq:Jeans}
   \frac{d(\rho \sigma_{r}^{2})}{dr} 
   = -\rho\frac{d\Phi}{dr} \ . 
\end{equation}
This yielded
\begin{equation}
\label{eq:initial-sigma}
  \sigma^{2}(r)=\frac{G}{\rho(r)}\int_{r}^{\infty} 
    \frac{M(r')\rho(r')}{r'^{2}}dr' \ . 
\end{equation}
To increase the amount of violent relaxation in the simulations, we
set the gravitational constant to $G=0.5$ in this calculation instead
of the $G=1.0$ that was used in the simulation. This ensured that each
halo would simultaneously collapse upon itself while merging with the
other haloes. 

\label{sec:init-test}
\begin{table}
 \label{tab:V0}
 \centering
  \begin{tabular}{@{}ll}
  \hline
   $V_0(r_0)$       &   $f_0$  \\ 
   \hline\hline
   $V_0(0.30)$ & 0.495881\\
   $V_0(0.33)$ & 0.416877\\  
   $V_0(0.37)$ & 0.337873\\  
   $V_0(0.43)$ & 0.258869\\  
   $V_0(0.53)$ & 0.179865\\  
   $V_0(0.71)$ & 0.107462\\  
   $V_0(0.73)$ & 0.100861\\  
   $V_0(0.96)$ & 0.060260\\  
   $V_0(1.16)$ & 0.041870\\  
   $V_0(1.32)$ & 0.032079\\  
   $V_0(1.47)$ & 0.026000\\  
   $V_0(1.60)$ & 0.021858\\
\hline
\end{tabular}
\caption{The phase-space density in the 12 initial volumes
  $V_0(r_0)$. Each initial volume $V_0(r_0)$ is a union of 4 shells around the
  4 haloes. The outer radius of the shell is $r_0$ and the inner
  radius is $r_0/2$. The outer radius $r_0$ has values in the range $0.30 < r_0<1.60$}
\end{table}

\subsubsection{Test particles}
We used 12 initial $V_0$ regions.  Each region was defined as the
union of four identical thick shells in \emph{real} space around each
one of the 4 haloes. Each shell extended from the radius $r_0/2$ to
$r_0$, with $r_0$ taking 12 possible values from 0.3 to 1.6. We denote
the $V_0$ region that corresponds to $r_0$ by $V_0(r_0)$. The initial
values of $r_0$ and the corresponding $f_0$ for each $V_0$ are given
in Table.~\ref{tab:V0}.

We used 100,000 test particles to sample each $V_0$ region by placing
25,000 particles in every shell.  The test particles were put in
locations where the initial phase-space density \Eq{eq:initial-f} was
exactly $f_0=(2\pi)^{-3/2}\rho(r_0)/\sigma^{3}(r_0)$.  The position
and velocity of a test particle were chosen in the following way:
first we chose the particle position by randomly picking a location
within the shell, in such a way that the real-space distribution of
test particles would be homogeneous and isotropic. Once $\Br$ was set
it implied a value for $|\Bv|$ by requiring that the RHS of
\Eq{eq:initial-f} would be equal to $f_0$.  The directionality of
$\Bv$ was then set isotropically.

Strictly speaking, one should add the contributions to $f(\Br,\Bv)$
from the three other haloes. However, by taking $r_0$ small enough we
have ensured that these contributions would never be larger than 10\%
of $f_0$.

\subsection{The $S_1$ surfaces}

The $S_1$ surfaces were chosen as two-dimensional shells in real space
whose centres were at the centre of the equilibrium halo with radii of
$r_1=0.3, 0.4, 0.5, 0.6, 0.7$.  In accordance with \Sec{sec:num-exp},
the measurements were done well within the centre of the relaxed halo
where the dynamical time is short and violent relaxation is believed
to be efficient, see Fig.~\ref{fig:final-profile} and \Sec{sec:n-body}
for more details.  To specify a particular $S_1$ surface we use the
notation $S_1(r_1)$.  We define the centre of the halo by calculating
its centre of mass using an iterative centre of mass (COM) approach.

The velocity coordinates of the $S_1$ surfaces were chosen to be the
velocity of the centre of mass, which amounts to $\Bv=0$ in the
centre of mass frame. The underlying assumption here is that the
equilibrium phase-space density of the test particles is isotropic
when measured in that frame, and therefore for $\bar{\tau}=(\Br,\Bv)$
the \emph{un-averaged} (i.e., local) conditional probability is given
by 
\begin{equation}
  K(\bar{\tau}, \tau) = 
    K\big(|\Br|, |\Bv|, (\Br\cdot\Bv),\tau\big) \ . 
\end{equation}
Such a function is constant over surfaces with $|\Br|=r_1$ and
$\Bv=0$.  Support for this assumption is found in the Poissonian
errors in $\left<K(S_1, V_0)\right>$ that we measured using the DTFE
and \Eq{eq:DTFE-average}. In all the measurements we obtained an error
$\lesssim 30\%$ - which is the typical Poissonian error that was found
when measuring the phase-space density in the spherical and isotropic
Hernquist-like distribution in
Eqs.~(\ref{eq:initial-f}-\ref{eq:initial-sigma}) (see
\Fig{fig:check-DTFE} and \Sec{sec:esti}). If the phase-space density
on these surfaces was not constant we would have obtained much larger
errors.

\subsection{$N$-body simulations and numerical effects}
\label{sec:n-body}

All simulations were run using the parallel version of the publicly
available tree-code GADGET \citep{ref:Spr01}. The simulations
were run on COSMOS, a shared-memory Altix 3700 with 152 1.3-GHz
Itanium2 processors hosted at the Department of Applied Mathematics
and Theoretical Physics (Cambridge). 

The accuracy of a collisionless GADGET simulation can be determined by
three parameters: the gravitational softening $\epsilon$, the internal
time-step accuracy $\eta$ and the cell-opening accuracy parameter of
the force calculation $\alpha$. After a number of test simulations we
settled for the following values for the simulation parameters.  We
used a physically fixed force softening of $\epsilon=0.02$ for both
the massive and massless test particles in the simulations.  The
half-mass radius $r_{m/2}$ of the final collapsed halo was $\sim 2.7$
length units and hence the employed softening was $\epsilon=7\times
10^{-3} r_{m/2}$.  We chose timesteps according to $\Delta
t=\sqrt{2\eta \epsilon/ |\textbf{a} |}$ (time-step criterion 0 of
GADGET).  Here $\textbf{a}$ is the acceleration and $\eta$ is the
time-step accuracy parameter which we set to $\eta=0.01$. In the force
calculation we employed the new GADGET cell opening criterion with a
high force accuracy of $\alpha=0.001$, for details see
\cite{ref:Spr01}.

\begin{figure}
  \center{\includegraphics[scale=1.0]{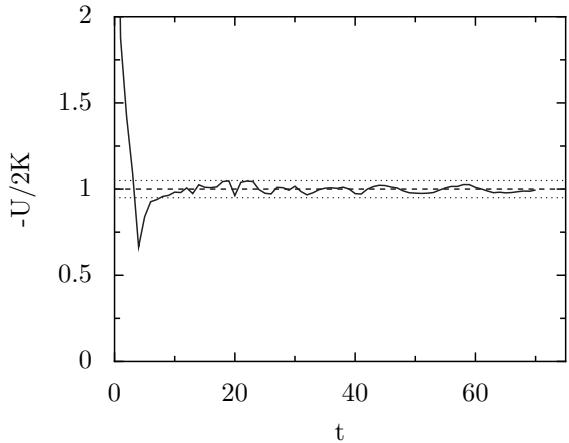}}
  \caption{The virial ratio $-U/2K$ as a function of time for one of
    the simulations. $U$ is the total potential energy and $K$ is the
    total kinetic energy.  Other simulations give very similar
    results. The dashed line denotes a ratio of 1 - for a fully
    relaxed system. For $t>30$ the virial ratio is always between
    $0.95$ and $1.05$, which are denoted by the dotted lines. The
    principal fluctuations occur at $t<20$, whereas at later times the
    gravitational potential is essentially constant and evolution
    proceeds through mixing.  }
  \label{fig:virial}
\end{figure}

\begin{figure}
  \center{\includegraphics[scale=1.0]{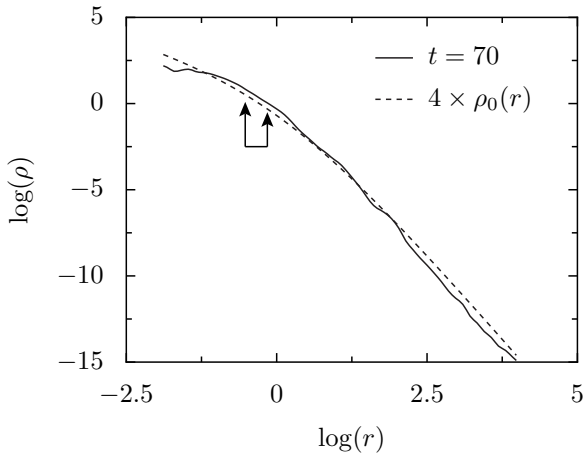}}
  \caption{The radial density profile of the relaxed system
    (solid line) at $t=70$. The dashed line is the initial radial
    density profile of the 4 haloes scaled by a factor of 4. The good
    fit between the two is in agreement with other numerical
    experiments such as \citet{ref:Boy04}. The two arrows denote the
    $r_1=0.3$ and $r_1=0.7$ radii which are the minimum and maximum
    radii at which the phase-space density of the test particles was
    measured. }
  \label{fig:final-profile}
\end{figure}

The simulations were ran for $t=70$ time units, with a typical run
taking $\sim 70,000$ timesteps to reach the final time.  In the final
snapshot, at least 70\% of the test particles were found within the
half-mass radius, the fraction rising to above 90\% for runs with low
values of $r_0$. At the half-mass radius the crossing time was
$t_{cross} \sim 10$, thus ensuring that the test particles that were
used for the measurement were fully relaxed. Further support for this
conclusion is presented in \Fig{fig:virial}, which shows the virial
ratio $-U/2K$ for one of the simulations. Here $U$ is the total
potential energy and $K$ is the total kinetic energy of the system.
For a fully relaxed system this ratio should be equal to one (the
virial theorem). We see that the system fluctuates strongly until
$t\sim 20$, after which the gravitational potential becomes constant
and the evolution of the phase-space density (both of the massive and
test particles) is through phase mixing only.
Figure~\ref{fig:final-profile} shows the radial density profile of the
massive particles at $t=70$. The dashed line is the function
$4\rho_0(r)$ with $\rho_0(r)$ being the initial Hernquist-like profile
of the four haloes given by \Eq{eq:initial-rho}. We see that the
profile of the relaxed halo is well fitted by the profile of the
initial haloes. This is in agreement with \citet{ref:Boy04} who found
the same behaviour in a head-on collision of two cuspy haloes.

An important test in assessing the reliability of a collisionless
simulation is to calculate to what extent the simulation is affected
by two-body relaxation. There have been a number of recent studies on
what regions can be considered reliable against two-body relaxation
[i.e. \citet{ref:Pow03, ref:Die04}]. We chose here to follow
\citet{ref:Boy04} who construct the local two-body relaxation time
$t_{r}$ as defined in terms of the circular velocity
$V_{circ}=V_{circ}(r)$, period $T(r)=2 \pi r/V_{circ}$ and the number
of particles $N$ (or equivalently mass $M$) interior to a radius r as
\begin{eqnarray}
   t_{r}(r) &=& T(r_{vir})\frac{N}{8 \ln N}
       \frac{r}{r_{vir}}\frac{V_{circ}}{V_{circ}(r_{vir})} \nonumber \\
         &=& \frac{\pi}{4}\frac{N}{\ln N}\sqrt{\frac{r^{3}}{GM(r)}}. 
\label{eq:trelax}
\end{eqnarray}
To minimise the effects of two-body relaxation we require that
$t_r(r)$ is longer than the overall simulation time for all values of
$r$ for which we measure the conditional probabilities.  We find for
the range $r=0.3-0.7$ for which the measurement of the final
phase-space density is done that the relaxation time is
$t_{r}=90-470$, thus concluding that the simulation is not affected
by two-body relaxation in our region of interest.

\section{Results }
\label{sec:results}

\begin{figure*}
 \begin{center}
   {\includegraphics[scale=0.7]{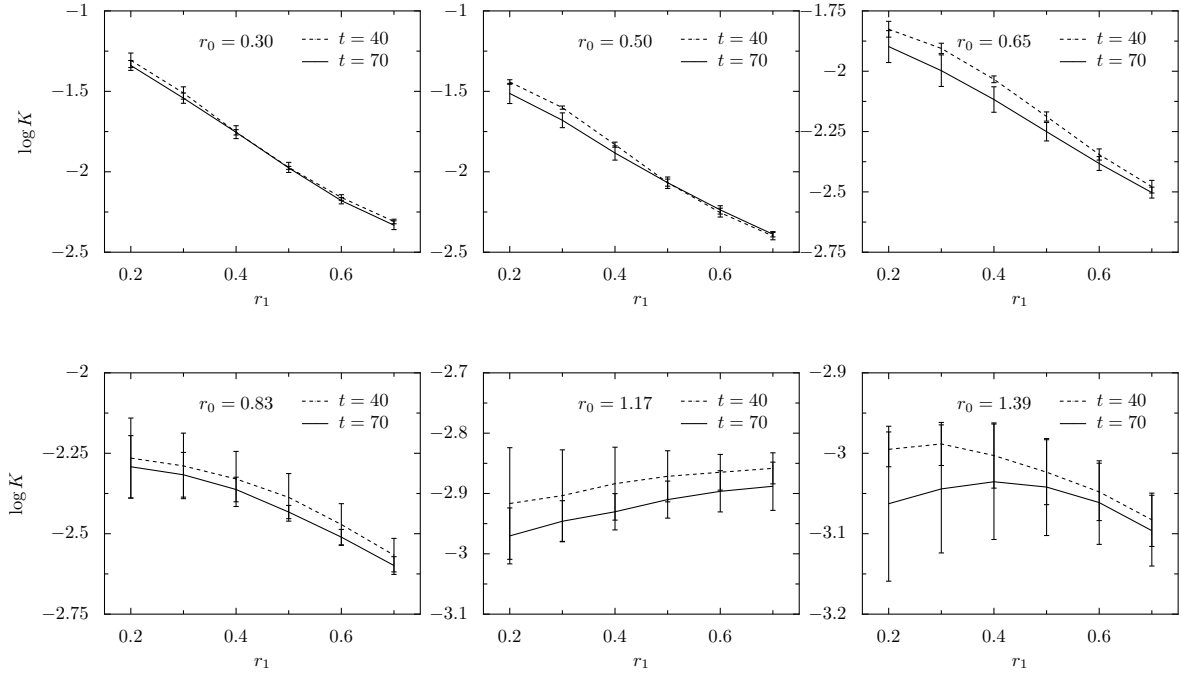}}
 \end{center}
 \caption[]{The average $\log K(r, r_0)$ for 6 out of the 12 possible
   values of the initial radius $r_0$. The dashed line is the average
   at $t=40$ whereas the smooth line is the average at $t=70$. The average was
   calculated using 5 different runs of the same simulation with
   different realisations of the initial conditions. The statistical
   error is typically smaller than $20\%$. The agreement between the
   $t=40$ and $t=70$ snapshots is also typically within an error of
   $20\%$. It is particularly good for the lowest values of $r_0$,
   indicating that these test particles are more relaxed that the ones
   with larger values of $r_0$.} 
 \label{fig:Ks}
\end{figure*}

Figure~\ref{fig:Ks} shows $\log \left<K[S_1(r_1), V_0(r_0)]\right>$ as
a function of $r_1$ for 6 out of the 12 initial conditions, at two
different times: $t_1=40$ time units and $t_2=70$ time units. To
simplify the notation, we denote it simply by $K(r_1, r_0)$. 

$K(r_1, r_0)$ is the result of averaging both over the $S_1(r_1)$
surfaces \emph{and} over the 5 different simulations that were run for
each initial region $V_0(r_0)$. We first calculated the $S_1$ average
and then calculated the average of the different realisations. Both
averages were done using $\log K$. In each realisation about 100-700
particles contributed to the $S_1(r_1)$ average. These particles
defined the Delaunay cells which were intersected by the 5D $S_1(r_1)$
surface.  The errors in \Fig{fig:Ks} are the scatter errors among the
different realisations.  They are typically in the range $20\%-30\%$,
and in general the two snapshots agree within that range.  We also
notice that for the lowest values of $r_0$ the agreement between the
two snapshots is better than for higher values of $r_0$.  This is an
indication that the phase-space structure of the low $r_0$ test
particles is more relaxed than that of the higher $r_0$. In addition,
for lower values of $r_0$, the spatial distribution is more
concentrated, having more test particles with smaller orbits. Such
particles have shorter dynamical times and thus tend to relax more
quickly than particles on outer orbits. It is therefore not surprising
that $K(r_1,r_0)$ with lower $r_0$ are less fluctuating than those
with higher values of $r_0$.

\begin{figure*}
 \begin{center}
   \includegraphics[scale=0.7]{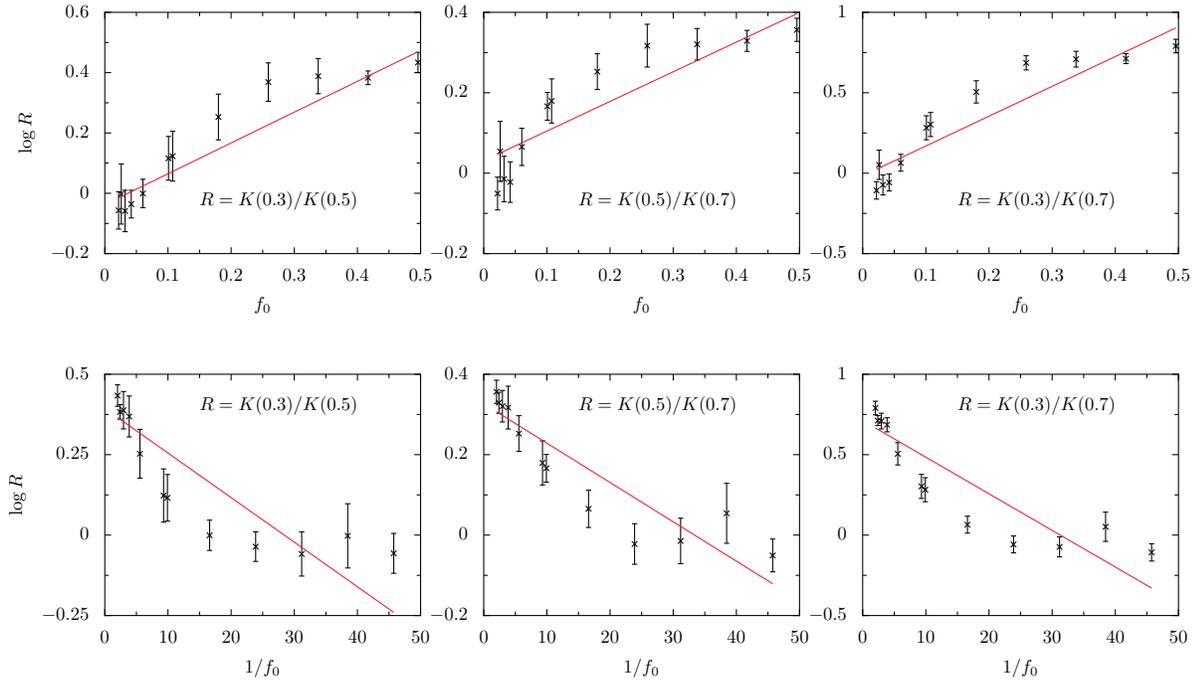}
 \end{center}
 \caption[]{The logarithm of the ratio $R(r_i, r_j)\EqDef
   K(r_i,r_0)/K(r_j, r_0)$ as a function of $f_0$ and of $1/f_0$, with
   $f_0$ being the initial phase-space density of the test
   particles. The $K(r_j,r_0)$ that was used to calculate the ratio is
   the average of 5 simulations, and is shown in \Fig{fig:Ks}. 

   For all plots a straight line was fitted using a least-mean-square
   procedure. The resultant $\chi^2$ for the three upper plots (LB67) is
   (left to right) $\chi^2=1.31, 2.51, 6.59$ and for the lower plots (NK00)
   $\chi^2=4.06, 2.43, 10.00$. All plots show a very strong
   non-linear behaviour, in contradiction with both LB67 and NK00
   theories.  }
 \label{fig:Rs}
\end{figure*}

Figure~\ref{fig:Rs} shows the log of the ratio $R(r_i, r_j) \EqDef
K(r_i, r_0)/K(r_j,r_0)$ calculated from the average $K(r,r_0)$ at
$t=70$. This is the main result of this paper. The upper plots show
$\log R$ as a function of $f_0$ for comparison with the LB67 theory,
whereas the lower plots show the same $\log R$ as a function of
$1/f_0$ for comparison with the NK00 theory. In both groups the pairs
$(r_i, r_j)$ are $(0.3,0.5)$, $(0.5,0.7)$ and $(0.3,0.7)$. Other
combinations of radii give similar results. 

In the upper plot we see that $\log R(r_i, r_j)$ increases until
$f_0\sim 0.25$ after which it becomes approximately constant. On the
other hand, in the lower plot $\log R(r_i, r_j)$ decreases until
$1/f_0\sim 25$, where again it saturates. Putting it all together we
see that $\log R(r_i, r_j)$ is approximately constant for $f_0<0.04$,
then it increases (as a function of $f_0$) until $f_0 \simeq 0.25$ and
then it saturates once more. $\log R(r_i, r_j)$ is therefore highly
non-linear both as a function of $f_0$ and as a function of $1/f_0$.
As was discussed in \Sec{sec:num-exp}, a linear behaviour is a
necessary condition for either theories to be correct.

To quantify the above departure from non-linearity, we fitted a
straight line for each of the size plots using a least-mean-square
procedure.  As expected, the results, are decisively against both
theories.  The quality of the fits is very bad, as can be clearly seen
from the plots. The reduced $\chi^2$ of the fits for the Lynden-Bell
theory are (left to right, top of Figure~\ref{fig:Rs})
$\chi_{LB}^2=1.31, 2.51, 6.59$ and for the Nakamura theory (left to
right, bottom of Figure~\ref{fig:Rs}) $\chi_{NK}^2=4.06, 2.43, 10.00$.
The lower $\chi^2$ for the first two fits is due to the small
differences between $r_i$ and $r_j$, which causes $R_{ij}$ to be very
close to unity and consequently $\log R_{ij}$ to approach zero, and
thus be better approximated by a straight line. On the other hand, in
the third pair where the difference between $r_i$ and $r_j$ is
maximal, $R_{ij}$ spans a wider range of values, and consequently both
theories produce a very poor fit of $\chi_{LB}^2=6.59$ and
$\chi_{NK}^2=10.00$.  This clearly shows that both theories fail
miserably the test of linearity of $\log R$ in $f_0$ (LB67) and
$1/f_0$ (NK00).

\section{Summary and conclusions}
\label{sec:conc}

We have performed a series of $N$-body simulations to test the
validity of the LB67 and NK00 theories.  Unfortunately, due to a
limited amount of time and computer resources we did not perform
simulations with different $f_0(\Br,\Bv)$ or with higher number of
particles. Different initial conditions and different resolutions may
yield qualitatively different equilibrium states [see, for example,
\citet{ref:Mer03}].  The running of additional simulations is left for
possible future work.  However, in light of the strong non-linearity
of the plots in \Fig{fig:Rs}, and the use of a very weak and general
condition to test the theories, we believe that our conclusions will
not be altered by additional simulations.

The main aspects in which our numerical experiment differ from
previous studies are:
\begin{itemize}
\item We used 3D simulations whereas previous studies concentrated on
  1D simulations. 
\item The experiment was explicitly designed to check which one of the
  two possible formulas of entropy is more correct: the LB67 formula
  which is derived using an equal-volume discretisation or the NK00
  formula which is derived using an equal-mass discretisation. 
\item To distinguish between the two theories we used a very weak
  condition on the conditional probabilities $K_i(\tau)$, which was
  estimated by measuring the phase-space densities of sets of test
  particles. Therefore we did not need the full analytical solution of
  the theories with all its computational and conceptual difficulties.
\item Previous attempts to verify the LB67 theory were done with the
  water-bucket initial conditions, in which the initial phase-space
  density has only one level. Here, as we needed to distinguish
  between the LB67 and NK00 theories, the initial phase-space density
  covered a continuous range. 
\end{itemize}

The results of the experiment are summarised in \Fig{fig:Rs}. They
provide very strong evidence against the LB67 and NK00 theories. As
was discussed in \Sec{sec:num-exp}, the linearity of $\log R$ in $f_0$
or $1/f_0$ is a very basic and weak requirement of both theories.
Therefore the non-linearity of all the plots in \Fig{fig:Rs} must be
attributed to the failure of the most basic assumption in these
theories, i.e., the maximisation of entropy.  Indeed, we cannot
explain the failure of the theories by the existence of additional
conserved quantities that depend on the $f(\Br,\Bv)$ (such as the
total angular momentum vector $\B{L}$) since we have used the final
$f(\Br,\Bv)$ itself as a constraint. We also cannot argue that our
measurements reflect the incomplete relaxation of the outer parts of
the system since they are done in the innermost parts of the system
(less than $1/2 r_{m/2}$) and in addition the linearity condition is
independent of the \emph{full} solution of the theories that assumes
complete relaxation in \emph{all} parts of the system. As was
explained in \Sec{sec:num-exp}, the weaker the condition is, the
deeper is the failure of the theories if they do not pass it - and
this is the bottom line.

In some respect this is a disappointing result as it shows that
neither of the theories is even remotely correct. Additionally we are
unable to decide which definition of the entropy is the ``right'' one
as they both seem to perform equally bad.

On the other hand, one may find some sort of comfort in the fact that
\emph{both} theories fail, as there is no solid a priori theoretical
argument against either equal-volume discretisation or equal-mass
discretisation, and it is not obvious why they should produce such
different results in the first place. Additionally, as was shown
recently by \citet{Ara05}, both theories contain some sort of
self-inconsistency since they are both non-transitive. Knowing that
the theories are fundamentally wrong empirically thus solves many of
these problems or at least makes them less relevant.

\section*{Acknowledgments}
We would like to thank D.~Lynden-Bell and S.~Colombi for very useful
discussions. I.A. would like to thank A.~Dekel and G.~Mamon for enabling
his stay at the IAP in Paris. 

This work was supported by a Marie-Curie Individual Fellowship of the
European Community No.  HPMF-CT-2002-01997.

The simulations were run on the COSMOS (SGI Altix 3700) supercomputer 
at the Department of Applied Mathematics and Theoretical Physics 
in Cambridge. COSMOS is a UK-CCC facility which is supported by 
HEFCE and PPARC.

\label{lastpage}

\end{document}